\documentclass[journal,draftcls,onecolumn,12pt,twoside]{IEEEtranTCOM}

\normalsize

\ifCLASSINFOpdf

\else

\fi

\usepackage{cite}
\usepackage{amsmath,amssymb,amsfonts}
\usepackage{algorithmic}
\usepackage{graphicx}
\usepackage{textcomp}
\usepackage{xcolor}
\usepackage{lipsum}
\usepackage{tikz}
\usepackage{amsthm}
\newtheorem{theorem}{Theorem}

\theoremstyle{definition}

\usepackage{commath}
\usepackage{mathtools}
\usepackage{tabularx}
\usepackage{multirow}
\usepackage{booktabs}
\usepackage{dcolumn}

\usepackage{array}

\graphicspath{ {./images/} }
\def\BibTeX{{\rm B\kern-.05em{\sc i\kern-.025em b}\kern-.08em
    T\kern-.1667em\lower.7ex\hbox{E}\kern-.125emX}}
\DeclarePairedDelimiterX\MeijerM[3]{\lparen}{\rparen}%
{\begin{smallmatrix}#1 \\ #2\end{smallmatrix}\delimsize\vert\,#3}

\newcommand\MeijerG[8][]{%
  G^{\,#2,#3}_{#4,#5}\MeijerM[#1]{#6}{#7}{#8}}
\begin{document}
\title{Performance Analysis for Reconfigurable Intelligent Surface Assisted MIMO Systems}
\author{Likun ~Sui,~Zihuai~Lin,~\IEEEmembership{Senior member,~IEEE,} Pei~Xiao,~\IEEEmembership{Senior member,~IEEE,} 
and~Branka~Vucetic,~\IEEEmembership{~Fellow,~IEEE}% <-this % stops a space
\thanks{Likun Sui, Zihuai Lin and Branka Vucetic are with the School 
  of Electrical and Information Engineering, University of Sydney, NSW, 2006 Australia e-mail: \{likun.sui,zihuai.lin,branka.vucetic\}@sydney.edu.au}

\thanks{Pei Xiao is with the Institute for Communication Systems (ICS), University of Surrey, UK (e-mail: p.xiao@surrey.ac.uk )}}
% \author{Likun ~Sui,~\IEEEmembership{Member,~IEEE,}
%         Zihuai~Lin,~\IEEEmembership{Senior member,~IEEE,}
%         Pei~Xiao,~\IEEEmembership{Senior member,~IEEE,}
%         and~Branka~Vucetic,~\IEEEmembership{~Fellow,~IEEE
%         }% <-this % stops a space
%         }
% \thanks{Likun Sui, Zihuai Lin and Branka Vucetic are with the School 
%  of Electrical and Information Engineering, University of Sydney, NSW, 2006 Australia e-mail: \{likun.sui,zihuai.lin,branka.vucetic\}@sydney.edu.au}% <-this % stops a space
%  \thanks{Pei Xiao is with the Institute for Communication Systems (ICS), University of Surrey, UK (e-mail: p.xiao@surrey.ac.uk ) }% <-this % stops a space
% %  \thanks{TCOM version based on Michael Shell's bare{\textunderscore}jrnl.tex version 1.3.}}

% \markboth{IEEE Transactions on Communications}%
% {draft paper}

\maketitle

\begin{abstract}
%\boldmath
This paper investigates the maximal achievable rate for a given average error probability and blocklength for the reconfigurable intelligent surface (RIS) assisted multiple-input and multiple-output (MIMO) system. The result consists of a finite blocklength channel coding achievability bound and a converse bound based on the Berry-Esseen theorem, the Mellin transform and the mutual information. Numerical evaluation shows fast speed of convergence to the maximal achievable rate as the blocklength increases and also proves that the channel variance is a sound measurement of the backoff from the maximal achievable rate due to finite blocklength.
\end{abstract}


\IEEEpeerreviewmaketitle

\section{Introduction}

\IEEEPARstart{R}{ecently}, there has been a prodigious increase in the demand for higher data rates in wireless communication networks due to the escalating number of mobile and IoT devices together with dramatically increased services \cite{leng2020implementation,IoT_FD,RF_energy1}. To this end, many candidate solutions have been proposed to deal with this demand, such as multiple-input multiple-output (MIMO) and millimeter-wave (mmWave)/TeraHertz (THz) communications \cite{cellular7,cellular8,cellular9,mm_Wave_MIMO_Sina,xu2021THzUAV}. These technologies offer significant data rate gains but have power and hardware cost limitations. Generally speaking, they can be regarded as a way to achieve higher data rates by altering transmitter and receiver features without influencing the propagation channel.\par
A possible approach to overcome the issues mentioned above lies in the use of the recently-developed reconfigurable intelligent surface (RIS), which consists of a massive array of scattering elements \cite{xu2022THzRIS,chu2022IoT_RIS,Hu2021CodedRIS}. Since the RIS is a passive device with low energy consumption and without self-interference, it is regarded as a better technology than the backscatter and Multi-input multi-output (MIMO) relay \cite{hu2021Backscatter,Chen2021MIMO,hu2019Ambc,xing2018ambc,MIMO_capacity,network_capacity,fountaincodes2014}. The array of elements can be configured by controllers to reflect radio waves towards arbitrary angles so that we can apply phase shifts and modify polarization\cite{b1}. Unlike existing relay technologies \cite{WRN,JNCC,RCRC,NC0,NC1}, RIS can turn the hostile propagation environment into a favorable one due to its unique properties ameliorate the signal quality at the receiver side without consuming additional power.\par
Most prior works have demonstrated the advantages of the RIS in terms of the bit-error-rate performance and cell coverage. In contrast, this paper takes a more fundamental information-theoretic perspective on the performance of RIS-assisted MIMO communication systems at the finite blocklength regime.\par
\textbf{Related Work:} In \cite{b2}, a broad mathematical framework of the RIS-assisted wireless communication system over Rayleigh fading channel was presented and then a theoretical upper bound was derived. Moreover, the authors presented the relationship between the received signal-to-noise ratio (SNR) and the number of reflecting elements, indicating that the received SNR grew considerably as the number of reflecting elements increased. Thus the reliable transmission over a noisy channel could be still accomplished at low SNRs with the support of the RIS elements. The authors of \cite{b3} investigated the coverage expansion achieved by the RIS-assisted wireless communication system over quasi-static flat Rayleigh fading channels. Furthermore, compared with both direct link and relay-assisted wireless communication systems, the SNR gain and the delay outage rate of the RIS were investigated. In \cite{b4}, the authors studied the RIS's placement optimization in a cellular network to maximize the cell coverage. They developed a coverage maximization algorithm (CMA) to obtain the optimal RIS's orientation distance. The authors of \cite{b5,b6,b7} focused on the RIS-assisted multiple-input single-output (MISO) wireless communication system, for which efficient algorithms, such as Lagrangian dual transform, active and passive beamforming, were studied to address the non-convex maximization problem of the weighted sum-rate that can be achieved by all groups. The authors of \cite{b8} statistically characterized the RIS-assisted wireless communication system under the premise that all cascaded fading channels between the transmitter, RIS and receiver follow the Rayleigh distribution. Furthermore, the closed form expression of theoretical outage probability was derived and the accuracy of their results was validated.\par

\textbf{Contribution:}
We use the Berry-Esseen theorem, mutual information and unconditional information variance as the fundamental mathematical basis to obtain the achievability and converse bounds for the maximal achievable rate $R$ given a fixed average error probability $\epsilon$ and blocklength $n$ for a RIS MIMO system. We consider the case when the channel state information (CSI) is unknown to the transmitter and hence we apply equal power allocation in our system. To derive the achievability bound, we use the Berry-Esseen theorem and some other inequalities and show the exact probability density function (PDF) of the channel output. In the converse counterpart, we combine the upper bound on the auxiliary channel, which is a product of $m$ copies of the PDF of Gamma distributed variables by the Mellin transform and Meijer G-function, and the upper bound of its output space by Lebesgue measure to derive our converse bound. Furthermore, to complete our achievability and converse bounds, we utilize different modulation schemes in our RIS MIMO system, and compare the performance for each modulation scheme mainly in two aspects. One is the required blocklength to achieve a certain level of the maximal achievable rate and the other is how the channel variance affects the convergence's speed to the maximal achievable rate. \par
% needed in second column of first page if using \IEEEpubid
%\IEEEpubidadjcol

\subsection{Notation}
The modulus, real portion, and imaginary part of a scalar complex number $y$ are denoted by $|y|$, $\Re\{y\}$ and $\Im\{y\}$, respectively.
A random vector is denoted by a bold capital letter, and its realization is denoted by a bold lowercase symbol.
The identity matrix of dimension $n\times n$ is denoted as $\mathbf{I}_n$. The Hermitian transposition of a matrix $\mathbf{Y}$ is denoted by the superscript $\mathbf{Y}^{H}$.
The trace of matrix of $\mathbf{Y}$ is represented by $tr(\mathbf{Y})$.
A complex Gaussian distribution with a mean of $\mu$ and a variance of $\sigma^2$ is denoted as $\mathcal{CN}(\mu,\sigma^2)$. 
%
% in particular, a complex Gaussian random variable $X\sim\mathcal{CN}(0,\sigma^2)$ with independent and identically distributed zero mean Gaussian real and imaginary components is circularly symmetric. 
The Frobenius norm of a matrix $\mathbf{Y}$ is $\norm{\mathbf{Y}}=\sqrt{tr(\mathbf{Y}\mathbf{Y}^{H})}$. %$\mathbb{R}_{+}$ stands for the nonnegative real line; in particular, $\mathbb{R}_{+}^{m}$ is the nonnegative orthant of the $m$-dimensional real Euclidean spaces. 
The nonnegative real line is denoted by $\mathbb{R}_{+}$, while the nonnegative orthant of the $m$-dimensional real Euclidean spaces is denoted by $\mathbb{R}_{+}^{m}$.
%
%$\log(\cdot)$ denotes the natural logarithm. 
$\mathbb{E}[\cdot]$ and $\mathbb{P}[\cdot]$ represent the statistical expectation and the probability of an event, respectively.

The remainder of this paper is structured as follows. The system model is described in Section  \ref{sys}, and the concept of a channel code is reviewed. The achievability bound for our system is derived in Section \ref{ach}. The converse bound for the RIS MIMO system under study is presented in Section \ref{con}. In Section \ref{num}, numerical findings are presented. Finally, Section \ref{conclu} draws the conclusion.

\section{System Model}\label{sys}
\begin{figure}[!t]
    \centering
    \includegraphics[width=4.5in]{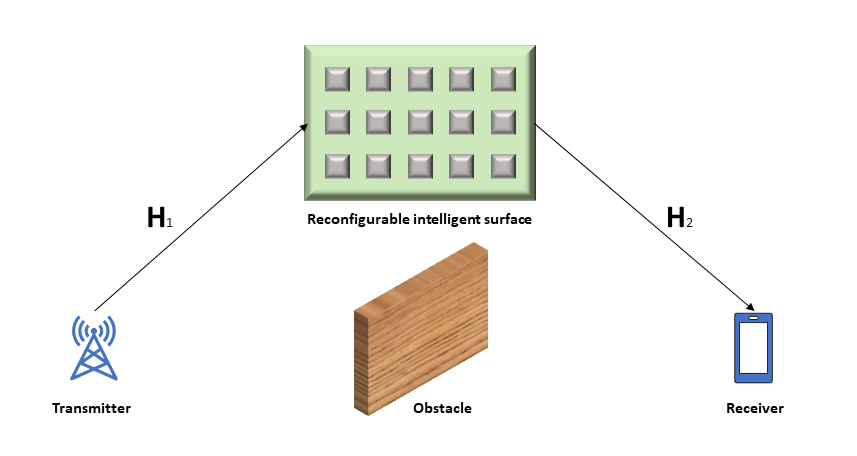}
    \caption{System Model.}
    \label{sys_model}
\end{figure}
We consider a RIS-assisted wireless communication system with $t$ transmit and $r$ receive antennas shown in Fig. \ref{sys_model}. Both of the transmitter and receiver have multiple antennas which are placed as uniform linear arrays (ULAs). The direct link is blocked by an obstacle (i.e. a wall or building) which is situated between the transmit antennas and the receive antennas. A rectangular RIS of $N_{ris}$ elements is utilized to improve the whole system performance, and only reflection-type RIS is considered in this paper. We assume that all the RIS elements are ideal which means that each of them can independently influence the phase and the reflection angle of the impinging wave.\par
We let $m=\min\{t,r\}$. The signal vector at the receive antenna array is given by
\begin{equation}
    \mathbf{Y} = \mathbf{H}\mathbf{X} + \mathbf{W},
\end{equation}
where $\mathbf{H}\in\mathbb{C}^{r\times t}$ is the channel matrix, $\mathbf{X}\in\mathbb{C}^{t\times n}$ is the transmit signal over $n$ channel uses, $\mathbf{Y}\in\mathbb{C}^{r\times n}$ is the corresponding received signal, and $\mathbf{W}\in\mathbb{C}^{r\times n}$ is the additive noise at the receiver, which is independent of $\mathbf{H}$ and has independent and identically distributed (i.i.d.) $\mathcal{CN}(0,1)$ entries. \par
% We assume that the total average transmit power has a maximum value of $P$, i.e., $\mathbb{E}\{\mathbf{X}^H\mathbf{X}\}\leq P$.
The channel matrix $\mathbf{H}$ of our RIS-assisted system can be expressed as
\begin{equation}\label{eqqq}
    \mathbf{H} = \mathbf{H}_2\boldsymbol\Sigma(\boldsymbol\theta)\mathbf{H}_1,
\end{equation}
where $\mathbf{H}_1\in \mathbb{C}^{N_{ris}\times t}$ represents the channel between the transmitter and the RIS, $\mathbf{H}_2\in\mathbb{C}^{r\times N_{ris}}$ represents the channel between the RIS and the receiver, and $\boldsymbol\Sigma(\boldsymbol\theta)=diag(\boldsymbol\theta)\in\mathbb{C}^{N_{ris}\times N_{ris}}$, where $\boldsymbol\theta=[\theta_1,\dots,\theta_{N_{ris}}]^{T}\in\mathbb{C}^{N_{ris}\times 1}$ represents the signal reflecting coefficient from the RIS. In this paper, similar to the related works \cite{b9,b10,b11}, we assume that the signal reflection from any RIS element is ideal, i.e., without any power loss. In other words, we may write $\theta_{i} = \exp\{j\phi_i\}$ for $i = 1,\dots,N_{ris}$, where $\phi_i$ is the phase shift induced by the $i$-th RIS element, which follows the uniform distribution in $[0,2\pi)$. Equivalently, we may write
\begin{equation}
    |\theta_i| = 1, \quad i = 1,\dots,N_{ris}.
\end{equation}\par
Throughout this paper, we define $\lambda_{max}(\cdot)$ as a function computing the $m$ largest eigenvalues of a channel matrix, and $\mathbf{g}= [g_1,\dots,g_m]^{T}$, then
\begin{equation}\label{eq5}
\mathbf{g} =\lambda_{max}\big(\mathbf{H}^H\mathbf{H}\big),  
\end{equation}
where $g_1\geq\dots\geq g_m$ are the $m$ largest eigenvalues.\par
Let us consider input and output sets $\mathcal{A}$ and $\mathcal{B}$ and a conditional probability measure $P_{\mathbf{Y}|\mathbf{X}}: \mathcal{A}\mapsto \mathcal{B}$. We denote a codebook with $M$ codewords by $(\mathbf{C}_1,\dots,\mathbf{C}_M)$. A decoder which can be defined as a random transformation $P_{Z|\mathbf{Y}}: \mathcal{B}\mapsto\{1,\dots,M\}$ which satisfies
\begin{equation}
    \frac{1}{M}\bigg(\sum_{i=1}^{M} P_{Z|\mathbf{X}}(i|\mathbf{C}_i)\bigg)=1-\epsilon,
\end{equation}
where $\epsilon$ is the average error probability. We also consider that each codeword $\mathbf{C}_i$ satisfies the equal power constraint $||\mathbf{C}_i||^2=nP$, where $P$ is the transmit power. Then, a codebook and a decoder whose average error probability is smaller than $\epsilon$ are termed as an $(n,M,\epsilon)$ code. In this paper, the information density also plays an essential role, which is defined as
\begin{equation}
    i(X;Y)=\log{\frac{dP_{XY}}{d(P_X\times P_Y)}(X,Y)}. 
\end{equation}\par
\section{Achievability Bound}\label{ach}

In this section, our achievability bound for the examined RIS MIMO system is presented below.
%Our achievability bound is based on the PPV bound \cite{b66}. 

\begin{theorem}\label{theorem1}
 %\par
% We consider that each codeword $\mathsf{X}$ satisfies the equal power constraint, and 

We consider a communication system having the finite input alphabet $\mathcal{A}$, and the continuous output alphabet $\mathcal{B}$. Let $p(\mathbf{Y},\mathbf{H}|\mathbf{X})$ be the corresponding conditional PDF on $\mathcal{B}$ for all $\mathbf{X}\in\mathcal{A}$, where $\mathbf{H}$ is a channel matrix which is distributed according to some density functions. The input distribution $P(\mathbf{X})\overset{\triangle}{=}[\mathbf{q}_0,\dots,\mathbf{q}_{t}]^T$, where $\mathbf{q}_i=[q_{i,0},\dots,q_{i,|\mathcal{A}|}]$ is equiprobable. Then we define the mutual information and the unconditional information variance as
\begin{equation}
    I(X;Y) \overset{\triangle}{=} \int_{0}^{\infty}\int_{-\infty}^{\infty}\sum_{\mathbf{X}\in\mathcal{A}^t}\bigg(P(\mathbf{X}) p(\mathbf{Y},\mathbf{H}|\mathbf{X})
    \log\big\{\frac{p(\mathbf{Y},\mathbf{H}|\mathbf{X})}{\sum_{\mathbf{X}'\in\mathcal{A}^t}P(\mathbf{X}')p(\mathbf{Y},\mathbf{H}|\mathbf{X}')}\big\}\bigg)d\mathbf{Y}d\mathbf{H},\label{eq777}
\end{equation}
\begin{multline}
    U(X;Y) \overset{\triangle}{=} \int_{0}^{\infty}\int_{-\infty}^{\infty}\sum_{\mathbf{X}\in\mathcal{A}^t}\bigg({P}(\mathbf{X}) p(\mathbf{Y},\mathbf{H}|\mathbf{X})
    \log^2\big\{\frac{p(\mathbf{Y},\mathbf{H}|\mathbf{X})}{\sum_{\mathbf{X}'\in\mathcal{A}^t}{P}(\mathbf{X}')p(\mathbf{Y},\mathbf{H}|\mathbf{X}')}\big\}\bigg)d\mathbf{Y}d\mathbf{H}\label{eq777b}\\-[I(X;Y)]^2.
\end{multline}
% Assuming that each $\mathbf{X}$ has the equal power constraint and belongs to the set:
%     \begin{equation}\label{eq1234}
%         F_n \overset{\Delta}{=} \{\mathbf{X}\in:\sum_{j=1}^m p_j(\mathbf{X})=P\}
%     \end{equation}\par
%     %For each codeword $\mathsf{X}\in F_n$, we have 
% with its corresponding power allocation vector, %which is
%     \begin{equation}\label{eq7}
%         \mathbf{p}(\mathbf{X})\in\mathbb{R}^{m}: p_j(\mathbf{X})=\frac{1}{n}\norm{\mathbf{x}_{j}}^2.
%     \end{equation}\par
Thus for the RIS MIMO channel and arbitrary $0<\epsilon<1$, we have the achievability bound 
\begin{equation}\label{eq44aa}
    R \geq I(X;Y)-\sqrt{\frac{U(X;Y)}{n}}Q^{-1}(\epsilon)+\frac{1}{n}+\mathcal{O}(n^{-\frac{3}{2}}),
\end{equation}
where $Q$ is the complementary Gaussian cumulative distribution function $Q(x) =\int_{x}^{\infty}\frac{1}{\sqrt{2\pi}}\\\exp(-\frac{u^2}{2})du$.
\end{theorem}
The proof of Th. \ref{theorem1} can be found below.\par

\begin{proof}

We need to introduce an important tool for proving Th. \ref{theorem1}, that is the Berry-Esseen theorem \cite{a5}.
\begin{theorem}\label{the1}[Berry-Esseen theorem]
    Let $X_k,k=1,\dots,n$ be independent with
    \begin{equation}\nonumber
        \mu_k=\mathbb{E}[X_k],\quad
        \sigma_k^2=Var[X_k],\quad
        t_k=\mathbb{E}[|X_k-\mu_k|^3],\quad
        \sigma^2=\sum_{k=1}^{n}\sigma_k^2 \quad \textrm{and} \quad
        T=\sum_{k=1}^{n}t_k.
    \end{equation}
    Then for any $-\infty<
    \tau<\infty$
    \begin{equation}
        \Big|\mathbb{P}\Big[\sum_{k=1}^{n}(X_k-\mu_k)\geq\tau\sigma\Big]-Q(\tau)\Big|\leq \frac{6T}{\sigma^3}.
    \end{equation}
\end{theorem}
For the proof of Th. \ref{theorem1}, we first need to prove that the second moment of $i(X;Y)$ is nonzero and its third moment is always less than infinite. 
\begin{align}
    &U(X;Y) = \mathbb{E}[|i(X;Y)-I(X;Y)|^2] \\\nonumber&= \int_{0}^{\infty}\int_{-\infty}^{\infty}\sum_{\mathbf{X}\in\mathcal{A}^t}\Big({P}(\mathbf{X}) p(\mathbf{Y},\mathbf{H}|\mathbf{X})(1-{P}(\mathbf{X}) p(\mathbf{Y},\mathbf{H}|\mathbf{X}))
    \log^2\big\{\frac{p(\mathbf{Y},\mathbf{H}|\mathbf{X})}{\sum_{\mathbf{X}'\in\mathcal{A}^t}{P}(\mathbf{X}')p(\mathbf{Y},\mathbf{H}|\mathbf{X}')}\big\}\Big)\\\nonumber&-2\binom{|\mathcal{A}|^t}{2}\bigg({P}(\mathbf{X}) p(\mathbf{Y},\mathbf{H}|\mathbf{X})
    \log\big\{\frac{p(\mathbf{Y},\mathbf{H}|\mathbf{X})}{\sum_{\mathbf{X}'\in\mathcal{A}^t}{P}(\mathbf{X}')p(\mathbf{Y},\mathbf{H}|\mathbf{X}')}\big\}\bigg)^2d\mathbf{Y}d\mathbf{H}\\\label{equ1}
    &=\int_{0}^{\infty}\int_{-\infty}^{\infty}\sum_{\mathbf{X}\in\mathcal{A}^t}\bigg({P}(\mathbf{X}) p(\mathbf{Y},\mathbf{H}|\mathbf{X})(1- p(\mathbf{Y},\mathbf{H}|\mathbf{X}))\\\nonumber&\quad   \log^2\big\{\frac{p(\mathbf{Y},\mathbf{H}|\mathbf{X})}{\sum_{\mathbf{X}'\in\mathcal{A}^t}{P}(\mathbf{X}')p(\mathbf{Y},\mathbf{H}|\mathbf{X}')}\big\}\bigg)d\mathbf{Y}d\mathbf{H}\\\label{equ2}
    &>0,
\end{align}
where (\ref{equ1}) follows from $2\binom{|\mathcal{A}|^t}{2}/|\mathbb{A}|^t=|\mathcal{A}|^t-1$ and ${P}(\mathbf{X})=1/|\mathcal{A}|^t$ and (\ref{equ2}) follows from $1- p(\mathbf{Y},\mathbf{H}|\mathbf{X})>0$.\par
Then, we need to show the third moment is less than infinite.
\begin{align}
    &T(X;Y)=\mathbb{E}[|i(X;Y)-I(X;Y)|^3]\\&=\Big|\mathbb{E}[|i(X;Y)|^3]+3I(X;Y)^2\mathbb{E}[|i(X;Y)|]-3I(X;Y)\mathbb{E}[|i(X;Y)|^2]-I(X;Y)^3\Big|\\
    &=\Big|\mathbb{E}[|i(X;Y)|^3]-3I(X;Y)\mathbb{E}[|i(X;Y)|^2]+2I(X;Y)^3\Big|\\
    &\leq \mathbb{E}[|i(X;Y)|^3]+2I(X;Y)^3\\\label{equ3}
    &\leq \mathbb{E}[|p(\mathbf{Y},\mathbf{H}|\mathbf{X})|^3]+\mathbb{E}[|\frac{1}{\sum_{\mathbf{X}'\in\mathcal{A}^t}{P}(\mathbf{X}')p(\mathbf{Y},\mathbf{H}|\mathbf{X}')}|^3]+2I(X;Y)^3\\\label{equ4}
    &\leq |\mathcal{B}|(3e^{-1}\log{e})^3+2I(X;Y)^3,
\end{align}
where (\ref{equ3}) follows from Holder's inequality and (\ref{equ4}) follows from $\max_{0<x<1}\{x\log^3{x}\}=0$ at $x=1$ and $\max_{0<x<1}\{x\log^3{\frac{1}{x}}\}=(3e^{-1}\log{e})^3$ at $x=e^{-3}$. We denote $i(X^n;Y^n)=\sum_{n}i(X;Y)$, and let its second moment $\sum_{n}U(X;Y)$ be nonzero and its third moment $\\\sum_{n}\mathbb{E}[|i(X;Y)-I(X;Y)|^3]<\infty$. Thus, Th.\ref{the1} is still applicable to $i(X^n;Y^n)$.\par
According to the DT bound in \cite{b66}, $\epsilon\leq\mathbb{E}\big[\exp\big\{-[i(X^n;Y^n)-\log \frac{M-1}{2})]^{+}\big\}\big]$, where $[\cdot]^{+}$ denotes $\max\{\cdot,0\}$. In the sequel, we prove that there exist some $\lambda$ values, so that
\begin{align}
     \epsilon &\geq \mathbb{E}\Big[\exp\{0\}1_{\{i(X^n;Y^n)-\log{\lambda}\leq 0\}}\Big]+\mathbb{E}\Big[\exp\big\{-i(X^n;Y^n)+\log\lambda\big\}1_{\{i(X^n;Y^n)-\log{\lambda}>0\}}\Big]\nonumber\\\label{eq1111}
     &=\mathbb{P}\bigg[i(X^n;Y^n)\leq\log{\lambda}\bigg]+
     \lambda\mathbb{E}\bigg[\exp\big\{-i(X^n;Y^n)\big\}1_{\{i(X^n;Y^n)>\log{\lambda}\}}\bigg].
\end{align}
The first step is to obtain the upper bound of the first part of the right-hand side of (\ref{eq1111}). After applying Th. \ref{the1}, we have
\begin{equation}\label{eq1111b}
    \mathbb{P}\bigg[i(X^n;Y^n)\leq nI(X;Y)-\tau\sqrt{nU(X;Y)}\bigg]\leq\frac{6T(X;Y)}{\sqrt{n}U(X;Y)^{\frac{3}{2}}}+Q(\tau).
\end{equation}
We assume 
\begin{equation}\label{eqbxx1}
    \log\lambda=nI(X;Y)-\tau\sqrt{nU(X;Y)},
\end{equation}
and
% And for large $n$, we denote 
% \begin{equation}\label{eq1111a}
%     \eta = Q^{-1}\bigg(\epsilon-\frac{2}{\sqrt{n}}\big(\frac{\log2}{\sqrt{2\pi}}+\frac{15T}{\sigma^2}\big)\bigg).
% \end{equation}
% Then substituting (\ref{eq1111a}) into (\ref{eq1111b}), we have
\begin{equation}\label{eq2222b}
    \mathbb{P}\bigg[i(X^n;Y^n)\leq\log{\lambda}\bigg]\leq \frac{6T(X;Y)}{\sqrt{n}U(X;Y)^{\frac{3}{2}}}+Q(\tau).
\end{equation}
The upper bound of the second part of the right-hand side of (\ref{eq1111}) is given below. For $0\leq i<\infty$ and any $\Delta>0$,
\begin{align}
    &\mathbb{P}\Big[-\sqrt{nU(X;Y)}(\tau-\frac{i\Delta}{\sqrt{nU(X;Y)}})\leq i(X^n;Y^n)\\\nonumber
    &\qquad\qquad\qquad\qquad\qquad-nI(X;Y)\leq -\sqrt{nU(X;Y)}(\tau-\frac{(i+1)\Delta}{\sqrt{nU(X;Y)}})\Big]\label{eq2222}\\
    &=\mathbb{P}\Big[\log\lambda+i\Delta\leq i(X^n;Y^n)\leq \log\lambda+(i+1)\Delta\Big]\\
    &\leq \frac{12T(X;Y)}{\sqrt{n}U(X;Y)^{\frac{3}{2}}}+Q(\tau+\frac{i\Delta}{\sqrt{nU(X;Y)}})-Q(\tau+\frac{(i+1)\Delta}{\sqrt{nU(X;Y)}}),
\end{align}
% \begin{align}
%     &\lambda\mathbb{E}\bigg[\exp\big\{-i(X^n;Y^n)\big\}1_{\{i(X;Y)>\log{\lambda}\}}\bigg]\\
%     &\leq \lambda\sum_{i=0}^{\infty}\exp\{-\log{\lambda}-i\delta\}
%     \mathbb{P}\bigg[\log{\lambda}+i\delta\leq i(X;Y)\leq \log{\lambda}+(i+1)\delta\bigg]\\
%     &\leq \lambda\sum_{i=0}^{\infty}\exp\{-\log{\lambda}-i\delta\}\bigg[Q(\frac{x}{\sigma})-Q(\frac{x+\delta}{\sigma})+\frac{12T}{\sigma^2}\bigg]\\
%     &\leq \frac{1}{\sigma}\bigg(\frac{\delta}{\sqrt{2\pi}}+\frac{12T}{\sigma^2}\bigg)\sum_{i=0}^{\infty}\exp\{-i\delta\},
% \end{align}
where (\ref{eq2222}) is obtained by applying Th. \ref{the1} twice. Then, 
% and setting $\delta=\log{2}$ and $1/\sigma=1/\sqrt{n}$, then we obtain
\begin{align}
    &\mathbb{E}\bigg[\exp\big\{-i(X^n;Y^n)\big\}1_{\{i(X^n;Y^n)>\log{\lambda}\}}\bigg]\\\label{eq2223}
    &=\sum_{i=0}^{\infty}\exp\{-(\log{\lambda}+i\Delta)\}\mathbb{P}\bigg[\log{\lambda}+i\Delta\leq i(X^n;Y^n)\leq \log{\lambda}+(i+1)\Delta\bigg]\\
    &\leq\sum_{i=0}^{\infty}\exp\{-(\log{\lambda}+i\Delta)\}\bigg[\frac{12T(X;Y)}{\sqrt{n}U(X;Y)^{\frac{3}{2}}}+Q(\tau+\frac{i\Delta}{\sqrt{nU(X;Y)}})-Q(\tau+\frac{(i+1)\Delta}{\sqrt{nU(X;Y)}})\bigg]\\\label{eq2224}
    &\leq \bigg(\frac{\Delta}{\sqrt{2\pi}\sqrt{nU(X;Y)}}+\frac{12T(X;Y)}{\sqrt{n}U(X;Y)^{\frac{3}{2}}}\bigg)\sum_{i=0}^{\infty}\exp\{-(\log{\lambda}+i\Delta)\},
\end{align}
where (\ref{eq2223}) is a result of the Riemann integral and (\ref{eq2224}) follows from the fact that for any $\sigma$, $Q(\frac{x}{\sigma})-Q(\frac{x+\Delta}{\sigma})\leq \frac{\Delta}{\sqrt{2\pi}\sigma}$. Thus, we have
\begin{align}\label{eq2222a}
    &\lambda\mathbb{E}\bigg[\exp\big\{-i(X^n;Y^n)\big\}1_{\{i(X^n;Y^n)>\log{\lambda}\}}\bigg]\\
    &\leq \lambda\bigg(\frac{\Delta}{\sqrt{2\pi}\sqrt{nU(X;Y)}}+\frac{12T(X;Y)}{\sqrt{n}U(X;Y)^{\frac{3}{2}}}\bigg)\sum_{i=0}^{\infty}\exp\{-(\log{\lambda}+i\Delta)\}\\
    &=\bigg(\frac{\Delta}{\sqrt{2\pi}\sqrt{nU(X;Y)}}+\frac{12T(X;Y)}{\sqrt{n}U(X;Y)^{\frac{3}{2}}}\bigg)\sum_{i=0}^{\infty}\exp\{-i\Delta\}\\\label{eq2225}
    &=\bigg(\frac{\Delta}{\sqrt{2\pi}\sqrt{nU(X;Y)}}+\frac{12T(X;Y)}{\sqrt{n}U(X;Y)^{\frac{3}{2}}}\bigg)\frac{\exp\{\Delta\}}{\exp\{\Delta\}-1},
\end{align}
where (\ref{eq2225}) follows for any $\exp\{x\}>1$, $\sum_{i=0}^{\infty}\exp\{-ix\}=\frac{\exp\{x\}}{\exp\{x\}-1}$.
Substituting (\ref{eq2222b}) and (\ref{eq2225}) into (\ref{eq1111}), we have
\begin{multline}\label{123}
    \mathbb{P}\bigg[i(X^n;Y^n)\leq\log{\lambda}\bigg]+\lambda\mathbb{E}\bigg[\exp\big\{-i(X^n;Y^n)\big\}1_{\{i(X^n;Y^n)>\log{\lambda}\}}\bigg]\\\leq Q(\tau)+\frac{1}{\sqrt{n}}\frac{6T(X;Y)}{U(X;Y)^{\frac{3}{2}}}\Big(1+2\frac{\exp\{\Delta\}}{\exp\{\Delta\}-1}+\frac{U(X;Y)\Delta\exp\{\Delta\}}{\sqrt{2\pi}6T(X;Y)(\exp\{\Delta\}-1)}\Big).
\end{multline}
Based on (\ref{eq1111}), we can assume that the right hand side of (\ref{123}) equals to $\epsilon$, then we obtain the value of $\tau$
\begin{equation}\label{eq2226}
    \tau = Q^{-1}\Big(\epsilon-\frac{1}{\sqrt{n}}\frac{6T(X;Y)}{U(X;Y)^{\frac{3}{2}}}\Big(1+2\frac{\exp\{\Delta\}}{\exp\{\Delta\}-1}+\frac{U(X;Y)\Delta\exp\{\Delta\}}{\sqrt{2\pi}6T(X;Y)(\exp\{\Delta\}-1)}\Big)\Big).
\end{equation}
For large $n$, the second item inside the Q function of (\ref{eq2226}) vanishes. Therefore, we can obtain $\tau=Q^{-1}(\epsilon)+\mathcal{O}(\frac{1}{\sqrt{n}})$. Then, we have $\log\lambda=nI(X;Y)-Q^{-1}(\epsilon)\sqrt{nU(X;Y)}+\mathcal{O}(\frac{1}{\sqrt{n}})$.\par
Thus,
\begin{align}
    &\log\frac{M-1}{2} \geq \log\lambda=nI(X;Y)-Q^{-1}(\epsilon)\sqrt{nU(X;Y)}+\mathcal{O}(\frac{1}{\sqrt{n}})\\
    &R \geq I(X;Y)-Q^{-1}(\epsilon)\sqrt{\frac{U(X;Y)}{n}}+\frac{1}{n}+\mathcal{O}(n^{-\frac{3}{2}}).
\end{align}
% \begin{equation}
%     \log{\lambda} = nI(X;Y)-\eta\sqrt{nU(X;Y)}).
% \end{equation}
% From (\ref{eq1111}), we can safely obtain
% \begin{equation}
%     \epsilon\geq \mathbb{E}\big[\exp\big\{-[i(X;Y)-\log\lambda]^{+}\big\}\big],
% \end{equation}
% where $[\cdot]^{+}$ denotes $\max\{\cdot,0\}$. According to the DT bound in \cite{b66}, we have
% \begin{equation}
%     \epsilon\leq\mathbb{E}\big[\exp\big\{-[i(X;Y)-\log (M-1)]^{+}\big\}\big].
% \end{equation}
% Then we complete the proof of Th. \ref{theorem1}
% \begin{equation}
%     R\geq \frac{\log{\lambda}}{n} = I(X;Y)-Q^{-1}(\epsilon)\sqrt{\frac{U(X;Y)}{n}}.
% \end{equation}
\end{proof}
To accomplish the achievability bound by applying Th. \ref{theorem1}, we need to obtain the exact expression of both (\ref{eq777}) and (\ref{eq777b}). At first, for our system model, the input distribution $P(\mathbf{X})=[\mathbf{q}_0,\dots,\mathbf{q}_{t}]^T$, where $\mathbf{q}_{i}=[\frac{1}{2},\frac{1}{2}]$ and $\mathbf{q}_{i}=[\frac{1}{4},\frac{1}{4},\frac{1}{4},\frac{1}{4}]$, for BPSK and QPSK respectively. And the conditional PDF of a MIMO Rayleigh fading channel, $p(\mathbf{Y},\mathbf{H}|\mathbf{X})$, is given by \cite{a3}\cite{a4}
\begin{equation}\label{eq444}
    p(\mathbf{Y},\mathbf{H}|\mathbf{X}) = p(\mathbf{H})p(\mathbf{Y}|\mathbf{X},\mathbf{H})=\frac{p(\mathbf{H})}{det(\pi \mathbf{I}_{r})}\exp\big(-(\mathbf{Y}-\mathbf{X}\mathbf{H})^{H}(\mathbf{Y}-\mathbf{X}\mathbf{H})\big),
\end{equation}
where $\mathbf{I}_{r}$ designates the $r\times r$ identity matrix and $det(\cdot)$ denotes the determinant.\par
Then
\begin{align}
    I(X;Y) &= \int_{0}^{\infty}\int_{-\infty}^{\infty}\sum_{\mathbf{X}\in\mathcal{A}^t}\bigg(P(\mathbf{X}) p(\mathbf{Y},\mathbf{H}|\mathbf{X})\log\big\{\frac{p(\mathbf{Y},\mathbf{H}|\mathbf{X})}{\sum_{\mathbf{X}'\in\mathcal{A}^t}P(\mathbf{X}')p(\mathbf{Y},\mathbf{H}|\mathbf{X}')}\big\}\bigg)d\mathbf{Y}d\mathbf{H}\\\label{eqq}
    &=\sum_{\mathbf{X}\in\mathcal{A}^t}\frac{q_{\mathbf{X}}}{det(\pi \mathbf{I}_r)}\underbrace{\int_{0}^{\infty}\cdots\int_{0}^{\infty}}_{\substack{\text{m-dimensions}\\\text{(w.r.t. $\mathbf{h}$)}}}\underbrace{\int_{-\infty}^{\infty}\cdots\int_{-\infty}^{\infty}}_{\substack{\text{m-dimensions}\\\text{(w.r.t. $\mathbf{y}$)}}}\prod_{i=0}^{m-1}p(\mathbf{h}_i)\exp\{-\frac{1}{2}||\mathbf{y}_i-\mathbf{h}_i\mathbf{x}_i||^2\}\\\nonumber
    &\bigg(-\log e\sum_{j=0}^{m-1}\frac{1}{2}||\mathbf{y}_j-\mathbf{h}_j\mathbf{x}_j||^2-\log\big\{\sum_{k=0}^{m-1}\sum_{\mathbf{x}_k'\in\mathcal{A}}\mathbf{q}_k\exp\{-\frac{1}{2}||\mathbf{y}_k-\mathbf{h}_k\mathbf{x}'_k||^2\}\big\}\bigg)d\mathbf{y}d\mathbf{h}.
\end{align}
and
\begin{align}
    U(X;Y) &= \int_{0}^{\infty}\int_{-\infty}^{\infty}\sum_{\mathbf{X}\in\mathcal{A}^t}\bigg(P(\mathbf{X}) p(\mathbf{Y},\mathbf{H}|\mathbf{X})\log\big\{\frac{p(\mathbf{Y},\mathbf{H}|\mathbf{X})}{\sum_{\mathbf{X}'\in\mathcal{A}^t}P(\mathbf{X}')p(\mathbf{Y},\mathbf{H}|\mathbf{X}')}\big\}\bigg)d\mathbf{Y}d\mathbf{H}\\\label{eqqq1}
    &=\sum_{\mathbf{X}\in\mathcal{A}^t}\frac{q_{\mathbf{X}}}{det(\pi \mathbf{I}_r)}\underbrace{\int_{0}^{\infty}\cdots\int_{0}^{\infty}}_{\substack{\text{m-dimensions}\\\text{(w.r.t. $\mathbf{h}$)}}}\underbrace{\int_{-\infty}^{\infty}\cdots\int_{-\infty}^{\infty}}_{\substack{\text{m-dimensions}\\\text{(w.r.t. $\mathbf{y}$)}}}\prod_{i=0}^{m-1}p(\mathbf{h}_i)\exp\{-\frac{1}{2}||\mathbf{y}_i-\mathbf{h}_i\mathbf{x}_i||^2\}\\\nonumber
    &\bigg(-\log e\sum_{j=0}^{m-1}\frac{1}{2}||\mathbf{y}_j-\mathbf{h}_j\mathbf{x}_j||^2-\log\big\{\sum_{k=0}^{m-1}\sum_{\mathbf{x}_k'\in\mathcal{A}}\mathbf{q}_k\exp\{-\frac{1}{2}||\mathbf{y}_k-\mathbf{h}_k\mathbf{x}'_k||^2\}\big\}\bigg)d\mathbf{y}d\mathbf{h}\\\nonumber
    &-[I(X;Y)]^2.
\end{align}
Next, we need to find the expression of $p(\mathbf{h}_i)$. According to (\ref{eqqq}), $p(\mathbf{h}_i)$ follows the Rayleigh distribution when $N_{ris}$ is sufficiently large. Thus,
\begin{equation}\label{eqqq2}
    p(\mathbf{h}_i) = \frac{2\mathbf{h}_i}{N_{ris}}\exp\{-\frac{\mathbf{h}^2_i}{N_{ris}}\}.
\end{equation}

Therefore, we can combine (\ref{eqq}), (\ref{eqqq1}) and (\ref{eqqq2}) together, and put the results into Th. \ref{theorem1}, then we can finally derive our achievability bound.

\section{Converse Bound}\label{con}

In this section, we derive the converse bound for the investigated RIS MIMO system on the basis of
%In this section, our converse bound is based on 
the meta-converse theorem \cite{b66} under the assumption of each codeword having an equal power. 
\begin{theorem}\label{theorem2}
    We consider the same equiprobable input distribution $P(\mathbf{X})$ and the same mutual information and unconditional information variance as defined in (\ref{eq777}) and (\ref{eq777b}), respectively, the converse bound for the RIS MIMO channel and arbitrary $0<\epsilon<1$ is given by, 
\begin{equation}\label{eq888}
    R \leq {I(X;Y)}-\sqrt{\frac{{U(X;Y)}}{n}}Q^{-1}(\epsilon+\frac{\epsilon}{\sqrt{n}})+\frac{(m+1)\log n}{2n}+\mathcal{O}(n^{-\frac{3}{2}}).
\end{equation}

\end{theorem}\par
The proof of (\ref{eq888}) can be found below.\par
\begin{proof}
We assume the transmitter is not aware of the realizations of the channel matrix $\mathbf{H}$. We denote the average power constraint 
\begin{equation}
    \mathbf{p}(\mathbf{X})\overset{\Delta}{=}\frac{1}{n}\mathbf{X}\mathbf{X}^H.
\end{equation}
Based on \cite{b17,b18,b19}, to evaluate the converse bound of an auxiliary channel, we need to obtain the lower bound of $\epsilon'$, which is the average error probability over the corresponding auxiliary channel. We thus denote the auxiliary channel $Q$ as:
\begin{equation}\label{eqq1}
    Q_{\mathbf{Y}|\mathbf{X},\mathbf{H}} \overset{\Delta}{=} \prod_{j=1}^{n}Q_{Y_{j}|\mathbf{X},\mathbf{H}},
\end{equation}
where
\begin{equation}
    Q_{Y_{j}|\mathbf{X},\mathbf{H}} = \mathcal{CN}(0,\mathbf{I}_r+\mathbf{H}\mathbf{p}(\mathbf{X})\mathbf{H}^H),
\end{equation}\par
We denote $\mathbf{B}\overset{\Delta}{=}\mathbf{I}_r+\mathbf{H}\mathbf{p}(\mathbf{X})\mathbf{H}^H$ and let its eigenvector  $\boldsymbol\omega= [\omega_1,\dots,\omega_m]=\lambda_{max}\big(\mathbf{B})$. Note that $\mathbf{P}=\mathbf{p}(\mathbf{X})$ is the only factor that affects the output of the $Q_{\mathbf{Y}|\mathbf{X},\mathbf{H}}$ channel. Let the space $\mathbf{S}\overset{\Delta}{=}\mathbf{p}(\mathbf{Y})=\frac{1}{n}\mathbf{Y}\mathbf{Y}^H$ and its entry is defined as the square of the norm of $\mathbf{Y}$ and is then normalized by the blocklength $n$, which is shown below %$Q_{S|\mathbf{p}}$ can be defined as below,
\begin{equation}\label{eq49}
    {S}_j = \frac{\omega_j}{n}\sum_{i=1}^{n}|Z_{j,i}|^2,\quad j=1,\dots,m,
\end{equation}
where $Z_{j,i}\sim\mathcal{CN}(0,1)$. $\mathbf{S}$ can be seen as the statistical expression of the receiver's detection of $\mathbf{X}$ from $(\mathbf{Y},\mathbf{H})$. Thus the auxiliary channel $Q_{\mathbf{Y}|\mathbf{X},\mathbf{H}}$ can be seen as $Q_{\mathbf{S}|\mathbf{B}}$. From (\ref{eq49}), we note that the $S_j$ follows the Gamma distribution, and its corresponding PDF is given by
\begin{equation}
    q_{S_j|B_j}(s_j|\omega_j) =
    \frac{n^n}{(\omega_j)^n\Gamma(n)}s_j^{n-1}\exp\Big\{-\frac{ns_j}{\omega_j}\Big\}.
\end{equation}\par
Moreover, as $Q_{\mathbf{S}|\mathbf{B}}$ is a product of $m$ copies of the PDF of $S_j$. We can obtain the PDF of $Q_{\mathbf{S}|\mathbf{B}}$ by the theorem shown below\cite{b28}.
\begin{theorem}\label{the2}
    Given $N$ independent Gamma-distributed random variables $x_i$ and that their shape parameter $k$ and scale parameter $\theta$ are all the same, we have the PDF of $x_i$ as
    \begin{equation}\label{eq40}
        f_i(x_i)=\frac{1}{\Gamma(k)\theta^k}x_i^{k-1}e^{-\frac{x_i}{\theta}}.
    \end{equation}
    We denote $z$ as the product of $N$ independent gamma variables $x_i$. Therefore, the PDF of $z=x_1x_2\dots x_N$ is a normalized Meijer G-function as
    \begin{equation}
        g(z)=\mathcal{K}\MeijerG[\big]{N}{0}{0}{N}{}{k-1}{\frac{z}{\theta^N}},
    \end{equation}
    where $\mathcal{K}$ is a normalizing factor which is 
    \begin{equation}
        \mathcal{K}=(\frac{1}{\theta})^N\prod_{i=1}^{N}\frac{1}{\Gamma(k)},
    \end{equation}
    and
    \begin{equation}
        \MeijerG[\big]{m}{n}{p}{q}{j_1,j_2, \dots, j_p}{k_1,k_2, \dots, k_q}{z}
        =\frac{1}{2\pi i}\int_{c-i^\infty}^{c+i^\infty}z^{-s}\cdot\frac{\prod_{j=1}^{m}\Gamma(s+k_j)\cdot\prod_{j=1}^{n}\Gamma(1-j_j-s)}{\prod_{j=n+1}^{p}\Gamma(s+j_j)\cdot\prod_{j=m+1}^{q}\Gamma(1-k_j-s)}ds,
    \end{equation}
    where $c$ is a vertical contour in the complex plane chosen to separate the poles of $\Gamma(s+k_j)$ from those of $\Gamma(1-j_k)-s$. 
\end{theorem}
The proof of Th. \ref{the2} can be found in Appendix \ref{C}.\par
We set two parameters, the shape parameter $k=n$ and the scale parameter $\theta_j=\frac{\omega_j}{n}$. The number of copies in our case is $N=m$. Then we can apply Th. \ref{the2} to calculate the PDF of $Q_{\mathbf{S}|\mathbf{B}}$ as
\begin{equation}\label{eq41}
    q_{S_j|B_j}(s_j|\omega_j)=\mathcal{K}\MeijerG[\big]{m}{0}{0}{m}{}{n-1}{s_j(\frac{n}{\omega_j})^m},
\end{equation}
where
\begin{equation}
    \mathcal{K}=(\frac{n}{\omega_j})^m\prod_{i=1}^{m}\frac{1}{\Gamma(n)},
\end{equation}
and
\begin{equation}
    \MeijerG[\big]{m}{0}{0}{m}{}{n-1}{s_j(\frac{n}{\omega_j})^m}
    =\frac{1}{2\pi i}\int_{c-i^\infty}^{c+i^\infty}(s_j(\frac{n}{1+\omega_j})^m)^{-z}\prod_{j=1}^{m}\Gamma(z+n-1)dz.
\end{equation}\par
Consider an arbitrary code for the auxiliary channel $Q$. The decoding sets corresponding to the $M$ codewords is denoted by $D_i,i=1,...,M$. $\epsilon'$ is the average error probability over the auxiliary channel $Q$. Then we have
\begin{align}
    1-\epsilon'&=\frac{1}{M}\mathbb{E}_{\mathbf{H}}\bigg[\sum_{i=0}^{M}\int_{D_i}q_{\mathbf{S}|\mathbf{B}}(\mathbf{s})d\mathbf{s}\bigg]\\
    &\leq \mathbb{E}_{\mathbf{H}}\bigg[\int_{D_0}q_{\mathbf{S}|\mathbf{B}}(\mathbf{s})d\mathbf{s}\bigg]\\
    &\leq \mathbb{E}_{\mathbf{H}}\bigg[\max\{q_{\mathbf{S}|\mathbf{B}}(\mathbf{s})\}\times Leb(D_0)\bigg].
\end{align}\par
Next we need to provide the upper bound of the output space of an arbitrary decoding set, $Leb(D_0)$. Due to the power allocation vector $\mathbf{p}(\mathbf{X})$, the space $\mathbf{P}$ can be bounded by a certain ball in $\mathbb{R}^{m}$. Based on the definition of $\mathbf{S}$, its space is a slightly larger ball than the space $\mathbf{P}$. Thus we can obtain the upper bounded Lebesgue measure \cite{b20} of $D_0$, 
\begin{equation}\label{eq17}
    Leb(D_0)\leq Leb(\mathbf{S}) \leq \frac{K}{M},
\end{equation}
where $Leb$ is the Lebesgue measure and $K$ is a constant.\par
% The proof of (\ref{eq17}) can be found in Appendix \ref{B}.\par
Then the decoding set of any codeword has a Lebesgue measure space which is always smaller than $\frac{K}{M}$. Therefore, we have
\begin{align}
    1-\epsilon'&\leq \mathbb{E}_{\mathbf{H}}\bigg[\max\{q_{\mathbf{S}|\mathbf{B}}(\mathbf{s})\}\times\frac{K}{M}\bigg]\\
    &\leq\frac{1}{M}\bigg(\frac{(n-1)^n\exp\{-(n-1)\}}{\Gamma(n)}\bigg)^m\mathbb{E}_{\mathbf{H}}\bigg[\prod_{i=1}^{m}\omega_j\bigg]\\\label{eq321}
    &=\frac{1}{M}\bigg(\frac{(n-1)^n\exp\{-(n-1)\}}{\Gamma(n)}\bigg)^m\times\int_{0}^{\infty}\prod_{i=1}^{m}(\omega_j)p(g)dg\\\label{eqq3}
    &\leq\frac{n^{m/2}}{M}.
\end{align}\par
According to the binary hypothesis testing in \cite{b66}, we have
\begin{align}
    \Lambda(\epsilon)&\geq\frac{1}{\lambda}\Big(\epsilon-\mathbb{P}\big[i(X^n;Y^n)\leq\log{\lambda}\big]\Big)\\\label{eqaxx}
    &\geq \frac{1}{\lambda}\Big(\epsilon-\frac{6T(X;Y)}{\sqrt{n}U(X;Y)^{\frac{3}{2}}}-Q(\tau)\Big),
\end{align}
where $\Lambda(\epsilon)$ denotes the average probability of error under $P_{\mathbf{Y}|\mathbf{X},\mathbf{H}}$ if the probability of error under $Q_{\mathbf{Y}|\mathbf{X},\mathbf{H}}$ is $\epsilon$ and (\ref{eqaxx}) follows from (\ref{eq2222b}).
Then,
\begin{align}
    \log\Lambda(\epsilon)&\geq -\log\lambda+\log\Big(\epsilon-\frac{6T(X;Y)}{\sqrt{n}U(X;Y)^{\frac{3}{2}}}-Q(\tau)\Big)\\\label{eqbxx}
    &=-n{I(X;Y)}+\tau\sqrt{n{U(X;Y)}}+\log\Big(\epsilon-\frac{6T(X;Y)}{\sqrt{n}U(X;Y)^{\frac{3}{2}}}-Q(\tau)\Big),
\end{align}
where (\ref{eqbxx}) follows from (\ref{eqbxx1}). We assume $\tau=Q^{-1}(\epsilon(1+\frac{1}{\sqrt{n}})-\frac{6T(X;Y)}{\sqrt{n}U(X;Y)^{\frac{3}{2}}})$. Thus,
\begin{equation}
     \log\Lambda(\epsilon)\geq-n{I(X;Y)}+\sqrt{n{U(X;Y)}}Q^{-1}(\epsilon(1+\frac{1}{\sqrt{n}})-\frac{6T(X;Y)}{\sqrt{n}U(X;Y)^{\frac{3}{2}}})-\frac{1}{2}\log n.
\end{equation}
Due to the fact that $\log\Lambda(\epsilon)\leq 1-\epsilon'$, we have
\begin{equation}\label{eqq2}
    -n{I(X;Y)}+\sqrt{n{U(X;Y)}}Q^{-1}(\epsilon+\frac{\epsilon}{\sqrt{n}})-\frac{1}{2}\log n+\mathcal{O}(\frac{1}{\sqrt{n}})\leq 1-\epsilon'.
\end{equation}\par
% Then we have
% \begin{equation}
%     \inf_{\mathsf{X}\in F_{n}}\log\beta_{1-\epsilon}(P_{\mathbf{Y}|\mathbf{X}},Q_{\mathbf{Y}|\mathbf{X}})= n\bigg(-\max_{X\in F_n}{I(X;Y)}-\sqrt{\frac{\min_{X\in F_n}{U(X;Y)}}{n}}Q^{-1}(\epsilon)-\frac{\log n}{2n}\bigg).
% \end{equation}
Thus substituting (\ref{eqq2}) into (\ref{eqq3}), we have
\begin{equation}
    R\leq {I(X;Y)}-\sqrt{\frac{U(X;Y)}{n}}Q^{-1}(\epsilon+\frac{\epsilon}{\sqrt{n}})+\frac{(m+1)\log n}{2n}+\mathcal{O}(n^{-\frac{3}{2}})
\end{equation}
This completes the proof.

\end{proof}
In order to complete the converse bound by applying Th. \ref{theorem2}, we use the same input distribution as in Section \ref{ach}. Then after we obtain the exact expression of $p(\mathbf{Y},\mathbf{H}|\mathbf{X})$ and $p(\mathbf{h}_i)$, we can combine (\ref{eqq}), (\ref{eqqq1}) and (\ref{eqqq2}) together, and put the results into Th. \ref{theorem2}, then we can finally derive the converse bound.\par
To compare with our result, we calculate the capacity of the channel whose input is a circularly symmetric complex Gaussian with zero mean and covariance $\frac{P}{t}\mathbf{I}_t$. The Theorem is shown below.
\begin{theorem}\cite{b12}
    % We consider the same channel with the same number of transmit and receive antennas as our system model under the power constraint $P$. Its capacity which is achieved by the complex Gaussian input equals
    Under the power constraint $P$, we assume the same channel with the same number of transmitting and receiving antennas as our system model. Its capacity, as determined by the complex Gaussian input, is equal to
\begin{equation}\label{eq666}
    \mathbb{E}_{\mathbf{g}}[\log(1+\frac{P}{t}g)]=\int_{0}^{\infty}\log(1+\frac{P}{t}g)dg,
\end{equation}
where according to (\ref{eq5}) in Section \ref{sys},
\begin{equation}\label{eq666aa}
    p(g) = \sum_{i=0}^m \frac{i!}{2(i+\max\{r,t\}-m)!}(L_{i}^{\max\{r,t\}-m}(g/N_{ris}))^2(g/N_{ris})^{\max\{r,t\}-m}\exp{(-g/N_{ris})}.
\end{equation}
Thus 
\begin{multline}
    C_{Gaussian} = \int_{0}^{\infty}\log(1+\frac{P}{t}g)\sum_{i=0}^m\frac{i!}{2(i+\max\{r,t\}-m)!}(L_{i}^{\max\{r,t\}-m}(g/N_{ris}))^2\\\times(g/N_{ris})^{\max\{r,t\}-m}\exp{(-g/N_{ris})}dg.
\end{multline}

\end{theorem}

\section{Numerical Results}\label{num}
\subsection{Evaluation of the Derived Bounds}
\begin{figure}[!ht]
    \centering
    \includegraphics[width=4.5in]{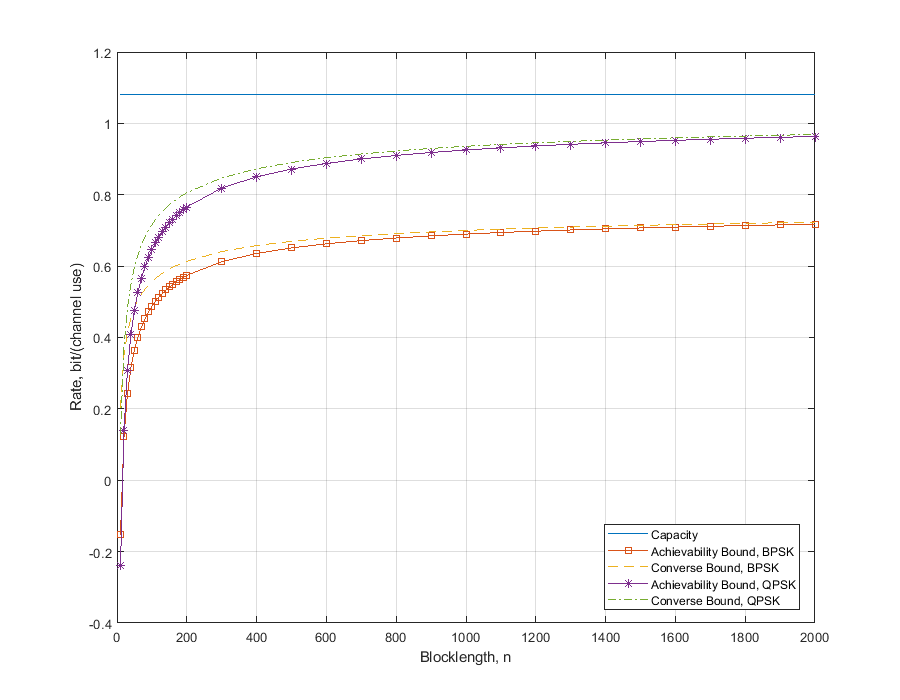}
    \caption{Achievability and converse bounds for $(n, M, \epsilon)$ codes for a RIS MIMO system over a Rayleigh fading channel and transmit antennas $t=2$ and receive antennas $r=1$, SNR=$-5$dB, $\epsilon=10^{-3}$, $N_{ris}=4$ and with BPSK and QPSK modulation, repectively.}
    \label{fig1}
\end{figure}

\begin{figure}[!ht]
    \centering
    \includegraphics[width=4.5in]{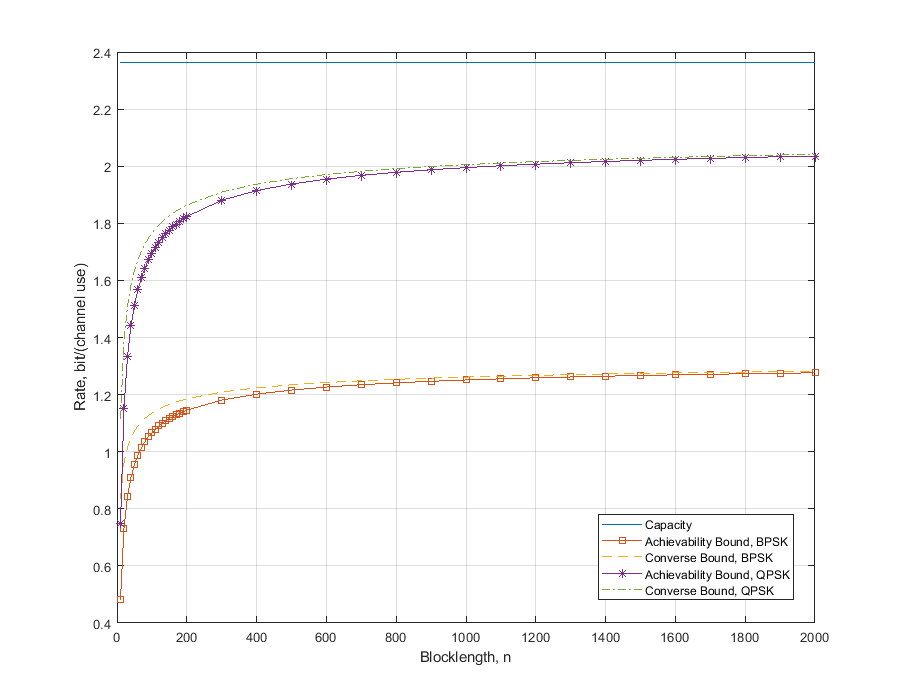}
    \caption{Achievability and converse bounds for $(n, M, \epsilon)$ codes for a RIS MIMO system over a Rayleigh fading channel and transmit antennas $t=2$ and receive antennas $r=1$, SNR=$-5$dB, $\epsilon=10^{-3}$, $N_{ris}=16$ and with BPSK and QPSK modulation, repectively.}
    \label{fig2}
\end{figure}

\begin{figure}[!ht]
    \centering
    \includegraphics[width=4.5in]{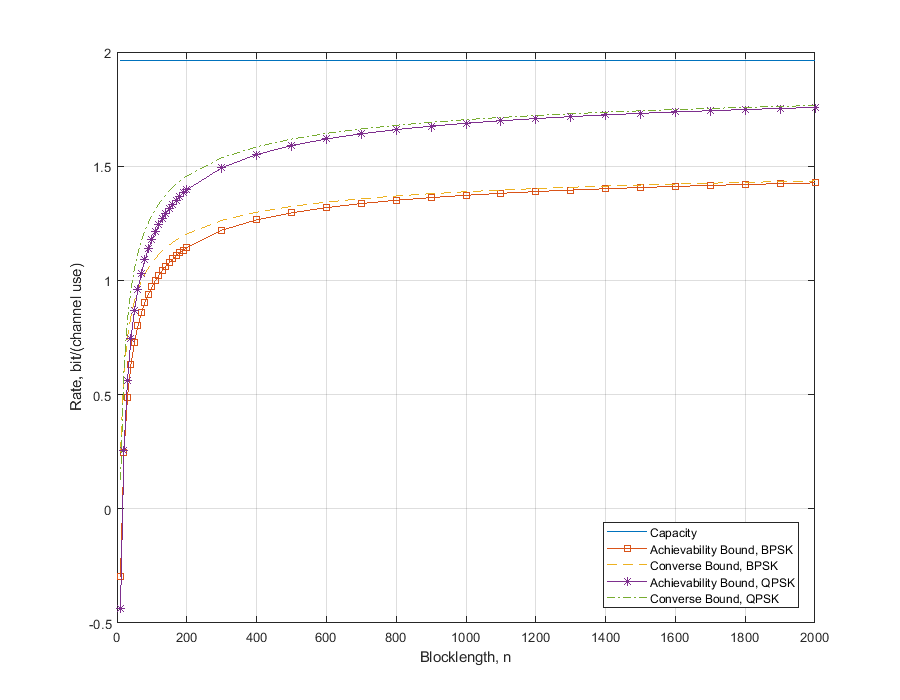}

    \caption{Achievability and converse bounds for $(n, M, \epsilon)$ codes for a RIS MIMO system over a Rayleigh fading channel and transmit antennas $t=2$ and receive antennas $r=2$, SNR=$-5$dB, $\epsilon=10^{-3}$, $N_{ris}=4$ and with BPSK and QPSK modulation, repectively.}
    \label{fig3}
\end{figure}
\begin{figure}[!ht]
    \centering
    \includegraphics[width=4.5in]{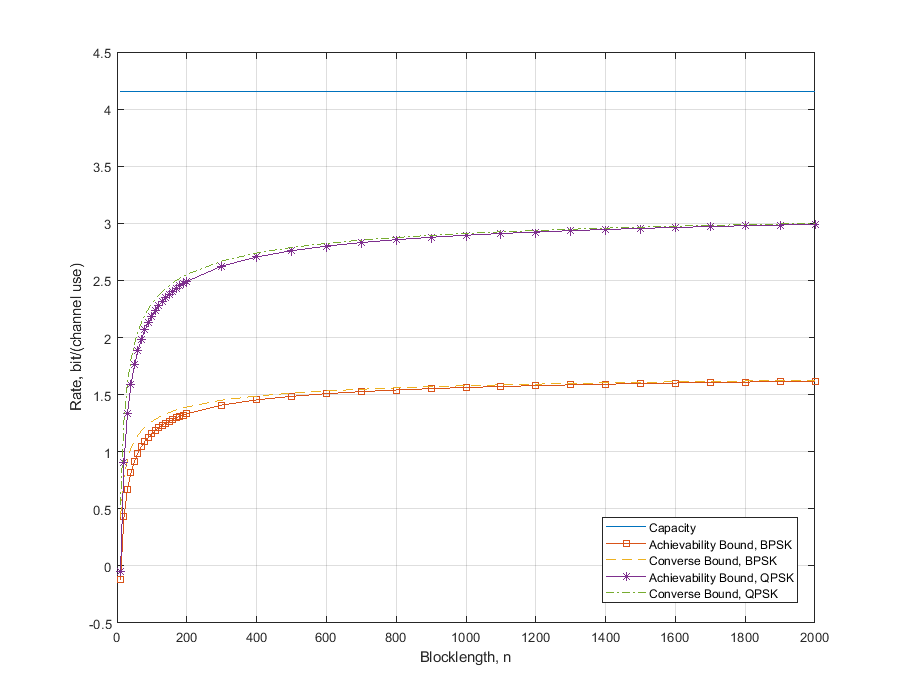}

    \caption{Achievability and converse bounds for $(n, M, \epsilon)$ codes for a RIS MIMO system over a Rayleigh fading channel and transmit antennas $t=2$ and receive antennas $r=2$, SNR=$-5$dB, $\epsilon=10^{-3}$, $N_{ris}=16$ and with BPSK and QPSK modulation, repectively.}
    \label{fig5}
\end{figure}

\begin{figure}[!ht]
    \centering
    \includegraphics[width=4.5in]{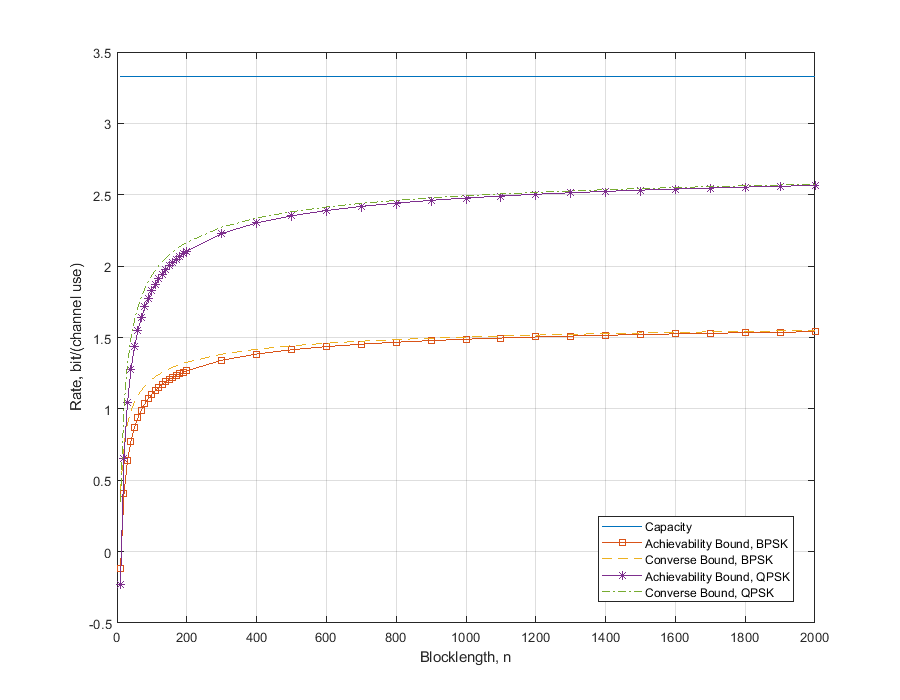}

    \caption{Achievability and converse bounds for $(n, M, \epsilon)$ codes for a RIS MIMO system over a Rayleigh fading channel and transmit antennas $t=2$ and receive antennas $r=2$, SNR=$-10$dB, $\epsilon=10^{-3}$, $N_{ris}=32$ and with BPSK and QPSK modulation, repectively.}
    \label{fig555}
\end{figure}
\begin{figure}[!ht]
    \centering
    \includegraphics[width=4.5in]{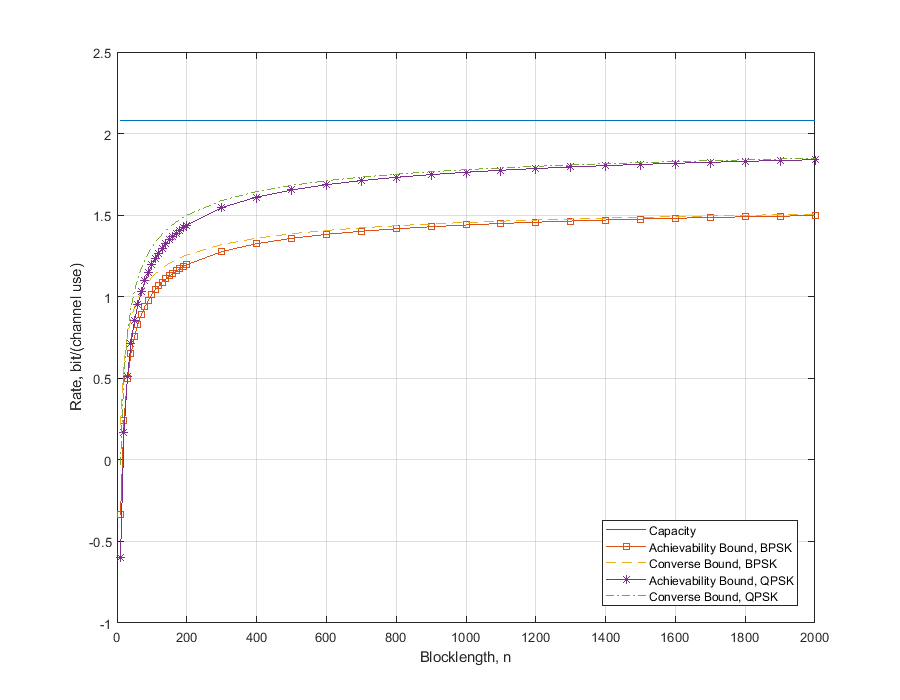}

    \caption{Achievability and converse bounds for $(n, M, \epsilon)$ codes for a RIS MIMO system over a Rayleigh fading channel and transmit antennas $t=3$ and receive antennas $r=2$, SNR=$-5$dB, $\epsilon=10^{-3}$, $N_{ris}=4$ and with BPSK and QPSK modulation, repectively.}
    \label{fig6}
\end{figure}
\begin{figure}[!ht]
    \centering
    \includegraphics[width=4.5in]{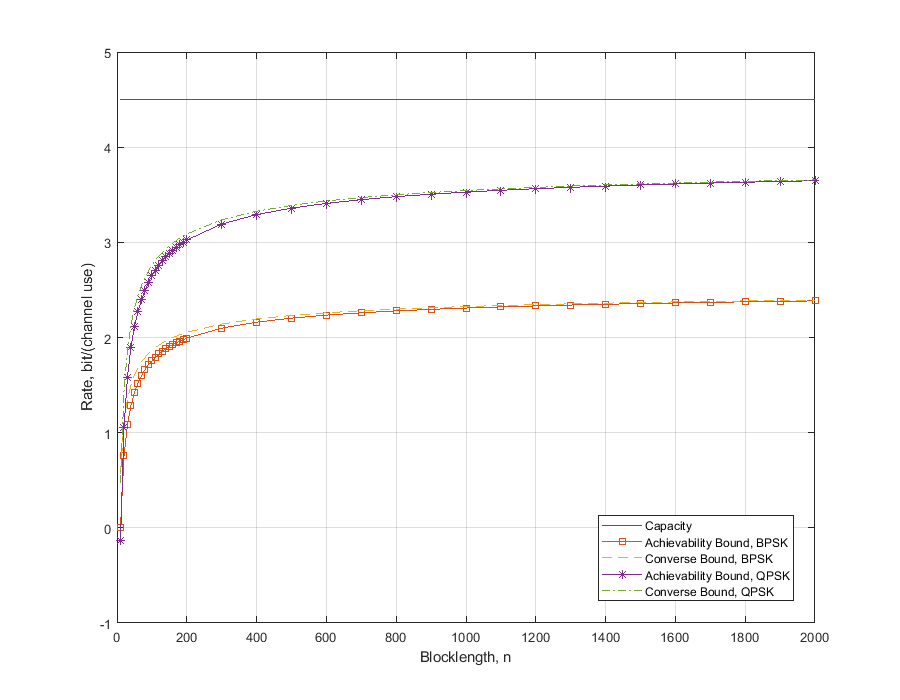}

    \caption{Achievability and converse bounds for $(n, M, \epsilon)$ codes for a RIS MIMO system over a Rayleigh fading channel and transmit antennas $t=3$ and receive antennas $r=2$, SNR=$-5$dB, $\epsilon=10^{-3}$, $N_{ris}=16$ and with BPSK and QPSK modulation, repectively.}
    \label{fig66}
\end{figure}

In this section, we consider a RIS MIMO system consisting of a transmitter with multiple transmitter antennas, a rectangular RIS of $N_{ris}$ elements and a receiver with multiple receive antennas. We assume all the channels, i.e., the channels between the transmitter and the RIS, the RIS and the receiver, the transmitter and the receiver, are independent with average error probability $\epsilon=10^{-3}$. Fig. \ref{fig1} shows the numerical results of the derived bounds with BPSK modulated and QPSK modulated signals and the capacity by assuming that all the channels are Rayleigh distributed, the numbers of transmit antennas and receive antennas are $t=2$, $r=1$, respectively and SNR=$-5$dB $N_{ris}=4$. From Fig. \ref{fig1}, we can see that $C_{Gaussian}=1.0811$ bit/(channel use), and the maximal achievable rate for BPSK modulation, which is calculated from (\ref{eq777}), is $0.7834$ bit/(channel use) and the blocklength $n$ required to achieve above $70\%$ and $80\%$ of its maximal achievable rate starts at $n=160$ and $n=360$, respectively. The gap between the capacity and its maximal achievable rate is $0.2977$ bit/(channel use). With the QPSK modulation, the maximal achievable rate, which is also obtained from (\ref{eq777}), is $1.0547$ bit/(channel use), and the blocklength $n$ required to achieve above $70\%$, and $80\%$ of its maximal achievable rate starts at $n=170$ and $n=380$, respectively. The gap in the QPSK case is $0.0264$ bit/(channel use).\par

In Fig. \ref{fig2}, we only change the RIS element from $N_{ris}=4$ to $N_{ris}=16$ and the rest parameters remain the same. The capacity, in this case, is $2.3629$ bit/(channel use). BPSK modulation's maximal achievable rate is $1.3367$ bit/(channel use). The blocklength $n$, which can surpass $70\%$ and $80\%$ of its maximal achievable rate, decreases dramatically to $50$ and $100$ compared with the case of $N_{ris}=4$. For $90\%$ of its maximal achievable rate, the required blocklength $n$ is $n=410$. Moreover, the gap increases to $1.0262$ bit/(channel use). For QPSK modulation, its maximal achievable rate is $2.1338$ bit/(channel use) and the blocklength $n=110$, and $n=420$ is required to achieve above $80\%$ and $90\%$ of its maximal achievable rate. The gap also enlarges from $0.0264$ bit/(channel use) to $0.2291$ bit/(channel use). From Fig. \ref{fig1} and \ref{fig2}, we can conclude that: 1) as $N_{ris}$ increases, the overall channel between the transmitter and the receiver becomes better. That means that the gap between the maximal achievable rate for different modulation schemes and the capacity increases and vice versa at the same SNR level. 2) the required blocklength $n$ falls significantly to achieve a given fraction of the maximal achievable rate as the number of the RIS elements increases.\par

The channel variance can be treated as the unconditional information variance (\ref{eq777b}). In the case of BPSK and QPSK modulation shown in Fig. \ref{fig1}, the channel variances are $0.9171$ and $1.7496$, respectively. In Fig. \ref{fig2}, the channel variance for BPSK and QPSK modulation is $0.7645$ and $2.0146$, respectively. It shows how quickly the performance converges to the maximum attainable rate as blocklength $n$ grows.
% It also implies that if the target transmits at a given fraction of the maximal achievable rate $0<\eta<1$ with a given average error probability $\epsilon$, the relationship between the required blocklength $n$ and the channel dispersion is given by
Additionally, if the target is to transmit at a fraction of the maximum achievable rate $0<\eta<1$ with an average error probability of $\epsilon$, the relationship between the required blocklength $n$ and the channel variance is as follows:
\begin{equation}
    n\approx\frac{U(X;Y)}{(I(X;Y))^2}\big(\frac{Q^{-1}(\epsilon)}{1-\eta}\big)^2.
\end{equation}\par
% \begin{table}[!ht]
% \caption{Required blocklength to achieve a given fraction of the maximal achievable rate for an AmBC MIMO system over a Rayleigh fading channel and transmit antennas $t=2$ and receive antennas $r=2$, $\epsilon=10^{-3}$, and $P(d)=[0.5,0.5]$.}
% \centering
% \begin{tabular}{l*{6}{D{.}{.}{-1}}}
% \toprule
%                     &\multicolumn{2}{c}{SNR$=-5$dB}                    &\multicolumn{2}{c}{SNR$=-10$dB}         \\\cmidrule(lr){2-3}\cmidrule(lr){4-5}
%                     &\multicolumn{1}{c}{BPSK}&\multicolumn{1}{c}{QPSK}&\multicolumn{1}{c}{BPSK}&\multicolumn{1}{c}{QPSK}\\
% \midrule
% Required $n$ to Achieve $80\%$ of The Maximal Achievable Rate&    $110$    &    $120$    &    $440$&     $470$\\
                    
% \addlinespace
% Required $n$ to Achieve $90\%$ of The Maximal Achievable Rate&    $450$    &    $500$    &    $1750$&    $1880$\\
                   
% \addlinespace

% \bottomrule
% \end{tabular}

% \label{table1}
% \end{table}
Figs. \ref{fig3}, \ref{fig5} and \ref{fig555} show the performance of the $2\times 2$ MIMO case. In Fig. \ref{fig3}, we only change the value of the receive antennas $r$ to $2$ and keep the rest of the parameters the same as in Fig. \ref{fig1}. The channel capacity is $1.9613$ bit/(channel use). The maximal achievable rates of BPSK and QPSK modulation are $1.5580$ bit/(channel use) and $1.9240$ bit/(channel use), respectively. The gap between the capacity of the $2\times 1$ MIMO and the $2\times 2$ MIMO cases is $0.8802$ bit/(channel use), which this gap is slightly lower than the $2\times 1$ MIMO's capacity in Fig. \ref{fig1} itself. Moreover, the gaps between the two maximal achievable rates and the channel capacity expands from $0.1769$ bit/(channel use) to $0.4033$ bit/(channel use) and $0.0106$ bit/(channel use) to $0.0373$ bit/(channel use) when comparing between the $2\times 1$ MIMO in Fig. \ref{fig1} and the $2\times 2$ MIMO in Fig. \ref{fig3}. To achieve $80\%$ of their maximal achievable rates, the required blocklengths for the two modulation schemes are $350$ and $380$, respectively. Referring to Fig. \ref{fig5}, we change the number of the RIS elements to $N_{ris}=16$ and keep the rest of the parameters unchanged. The capacity increases to $4.1535$ bit/(channel use). Compared with the case in Fig. \ref{fig2}, the capacity increases by $1.7906$ bit/(channel use). The respective maximal achievable rates are $1.7488$ bit/(channel use) and $3.2224$ bit/(channel use) for BPSK and QPSK modulation, respectively. The required blocklengths to achieve the same fraction above, which is $80\%$, are $n=280$ and $n=260$. In Fig. \ref{fig555}, we change the SNR to $-10$dB, and the number of the RIS elements to $N_{ris}=32$ and keep the rest of the parameters unchanged. The capacity in this case is $3.3262$ bit/(channel use). Furthermore, the maximal achievable rates of the two modulation schemes are $1.6666$ bit/(channel use) and $2.7776$ bit/(channel use). To reach $90\%$ of their maximal achievable rates, the required blocklengths are $n=1150$ and $n=1170$, respectively. Moreover, the channel variances for BPSK and QPSK scheme are $3.3402$ and $9.4568$, respectively.\par
In Figs. \ref{fig6}, \ref{fig66}, we demonstrate the performance of the $3\times 2$ MIMO case. For the combination of SNR$=-5$dB and $N_{ris}=4$, the capacity slightly increases from $1.9613$ bit/(channel use) to $2.0825$ bit/(channel use) compared with the $2\times 2$ MIMO case. Two maximal achievable rates for BPSK and QPSK increase to $1.6368$ bit/(channel use) and $2.0254$ bit/(channel use), respectively. The gaps of the maximal achievable rate between the $3\times 2$ MIMO and the $2\times 2$ MIMO cases are $0.0785$ bit/(channel use) and $0.1014$ bit/(channel use). Furthermore, the blocklengths which are needed to achieve $80\%$ of the maximal achievable rate increase from $350$ to $360$ for BPSK and from $380$ to $420$ for QPSK, respectively. In Fig. \ref{fig66}, we set the number of the RIS elements to $N_{ris}=16$. In terms of the capacity, the maximal achievable rates of the BPSK and the QPSK modulations, the gaps between the $3\times 2$ MIMO in Fig. \ref{fig66} and the $2\times 2$ MIMO in Fig. \ref{fig5} are $0.3495$ bit/(channel use), $0.8168$ bit/(channel use) and $0.7128$ bit/(channel use), respectively. To achieve $80\%$ of the $
3\times 2$ MIMO case, the required blocklengths are $n=250$ and $n=270$, respectively.\par
When we calculate $U(X,D;Y)$ in (\ref{eq777b}) for both of the modulation schemes, if $U(X,D;Y)=0$, then we need to replace unconditional information variance $U(X,D;Y)$ with conditional information variance $V(X,D;Y)$, which can be defined as
\begin{align}
    V(X,D;Y)\overset{\triangle}{=}& \mathbb{E}\big[Var(i(X,D;Y)|X)\big]\\
    =& \sum_{x\in\mathcal{A}}\bigg\{\int\sum_{d\in\mathcal{D}}p(y|x,d)\log^2\frac{p(y|x,d)}{p(y)}-\big[D\big(p(y|x,d)||p(y)\big)\big]^2\bigg\},
\end{align}
where $D(P||Q)$ denotes the divergence between distributions $P$ and $Q$.\par
\subsection{Rate vs SNR}
\begin{figure}[!ht]
    \centering
    \includegraphics[width=4.5in]{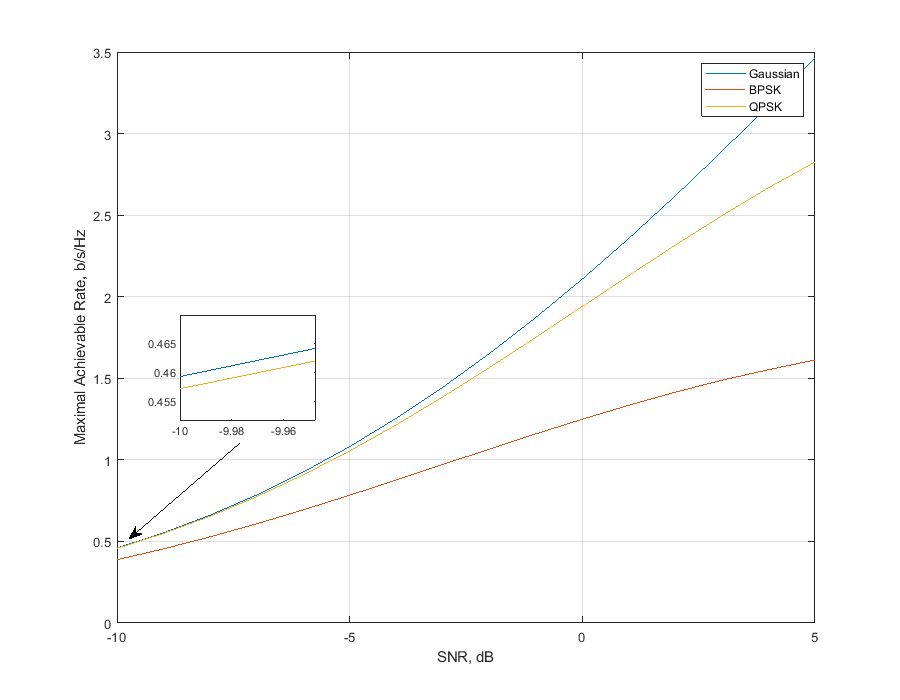}

    \caption{The maximal rate achieved by Gaussian inputs, QPSK, and BPSK in a RIS MIMO system over a Rayleigh fading channel and transmit antennas $t=2$ and receive antennas $r=1$, and $N_{ris}=4$}
    \label{fig8}
\end{figure}
\begin{figure}[!ht]
    \centering
    \includegraphics[width=4.5in]{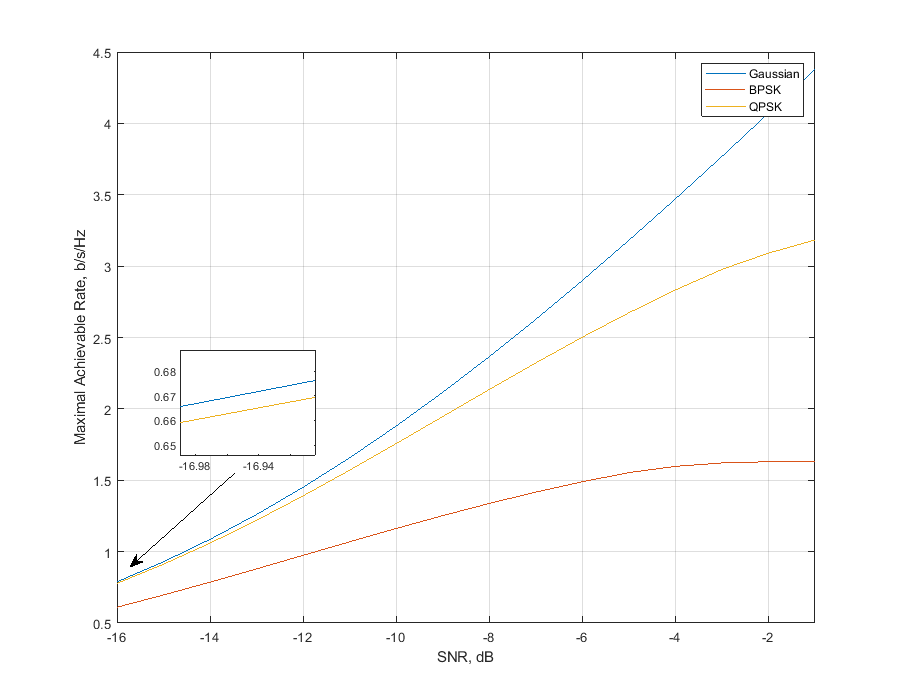}

    \caption{The maximal rate achieved by Gaussian inputs, QPSK, and BPSK in a RIS MIMO system over a Rayleigh fading channel and transmit antennas $t=2$ and receive antennas $r=1$, and $N_{ris}=32$}
    \label{fig9}
\end{figure}
\begin{figure}[!ht]
    \centering
    \includegraphics[width=4.5in]{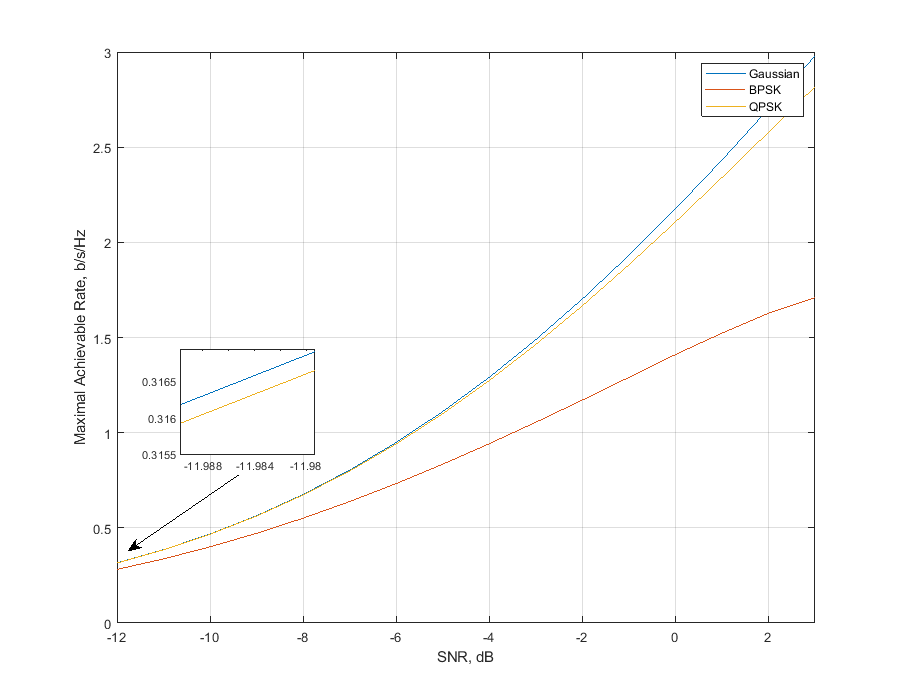}

    \caption{The maximal rate achieved by Gaussian inputs, QPSK, and BPSK in a RIS MIMO system over a Rayleigh fading channel and transmit antennas $t=3$ and receive antennas $r=1$, and $N_{ris}=4$}
    \label{fig10}
\end{figure}
In Figs. \ref{fig8} and \ref{fig9}, we illustrate the maximal achievable rates achieved by Gaussian inputs, QPSK and BPSK modulations in a RIS $2\times 1$ MIMO system with the number of the RIS elements $N_{ris}=4$ and $N_{ris}=32$, respectively. Fig. \ref{fig8} shows that the capacity of the channel achieved by circularly symmetric complex Gaussian inputs increases without any boundary as the SNR increases. However, the trends of the maximal achievable rates of each modulation scheme are similar to the Gaussian inputs at the low SNR regime. Then, the gaps between the Gaussian input and the QPSK modulated input, the Gaussian input and the BPSK modulated input increase as the SNR increases. At the high SNR regime, according to \cite{a6}, the upper bounds of the maximal rates achieved by BPSK and QPSK modulations go to $2$ bit/(channel use) for BPSK and $4$ bit/(channel use) for QPSK. In Fig. \ref{fig9}, the limits imposed by each modulation scheme are the same. However, the starting points move to smaller SNR values. When the number of the RIS elements increases, the channel condition becomes better and the required transmit power to achieve the same level of the rate decreases correspondingly. We change the number of the transmit antennas $t=3$ and $N_{ris}$ goes back to $4$. The result is shown in Fig. \ref{fig10}. At the same level of SNR$=2$dB, the maximal achievable rate of the QPSK modulation increases by $0.200$ bit/(channel use). The increasing trend in the $3\times 1$ MIMO case has not slowed down compared with the trend in Fig. \ref{fig8}. Moreover, the upper bounds of each modulation are $3$ bit/(channel use) for BPSK and $6$ bit/(channel use) for QPSK. \par
\section{Conclusion}\label{conclu}
In this paper, we have established achievability and converse bounds on the maximal achievable rate $R(n,\epsilon)$ at a given blocklength $n$ and an average error probability $\epsilon$ for a RIS MIMO system. The analytical results demonstrated that the number of transmit and receive antennas and the channel variance $U(X;Y)$ would affect the convergence speed to the maximal achievable rate as the blocklength $n$ increases. In this work, we only considered memoryless modulation schemes, for our future work, we will extend our work to memory modulation schemes, such as \cite{codedcpm1,codedcpm2,codedcpm3,codedcpm4,codedcpm5,codedcpm6}. 

% if have a single appendix:
%\appendix[Proof of the Zonklar Equations]
% or
%\appendix  % for no appendix heading
% do not use \section anymore after \appendix, only \section*
% is possibly needed

% use appendices with more than one appendix
% then use \section to start each appendix
% you must declare a \section before using any
% \subsection or using \label (\appendices by itself
% starts a section numbered zero.)
%

\appendices
\section{Proof of Th. \ref{the2}}\label{C}

    In this Appendix, we show the complete proof of Th. \ref{the2}. At first, we denote that the Mellin integral transform of $\exp{(-\frac{x}{\theta})}$ in (\ref{eq40}) is 
    \begin{equation}
        \mathcal{M}\{\exp{(-\frac{x}{\theta})}|s\}=\int_0^\infty x^{s-1}e^{(-\frac{x}{\theta})}dx
        =\theta^s\int_0^\infty x^{s-1}e^{-x}dx
        =\theta^s\Gamma(s),
    \end{equation}
    and
    \begin{equation}
        \mathcal{M}\{x^kf(x)|s\}=\mathcal{M}\{f(x)|s+k\}.
    \end{equation}
    Secondly, we derive that the Mellin transform of the PDF of the gamma variable in (\ref{eq40}) as
    \begin{equation}
        \mathcal{M}\{f_i(x_i)|s\}=\int_0^\infty x^{s-1} \frac{1}{\Gamma(k)\theta^k}x^{k-1}e^{-\frac{x}{\theta}}dx=\frac{\theta^{s-1}}{\Gamma(k)}\int_0^\infty x^{(k+s-1)-1}e^{-x}dx=\frac{\Gamma(k+s-1)}{\Gamma(k)}\theta^{s-1}.
    \end{equation}\par
    Thus we can obtain the Mellin transform of $g(z)$, which is defined as the PDF of the product of $N$ independent random Gamma variables, as
    \begin{equation}
        \mathcal{M}\{g(z)|s\}=\prod_{i=1}^{N}\mathcal{M}\{f_i(x_i)|s\}=\theta^{N(s-1)}\prod_{i=1}^{N}\frac{\Gamma(k+s-1)}{\Gamma(k)},
    \end{equation}
    and
    \begin{align}
        g(z)&=\frac{1}{2\pi i}\int_{c-i^\infty}^{c+i^\infty} z^{-s}\theta^{N(s-1)}\prod_{i=1}^{N}\frac{\Gamma(k+s-1)}{\Gamma(k)}ds\nonumber\\
        &=\frac{1}{2\pi i\theta^N}\int_{c-i^\infty}^{c+i^\infty} (\frac{z}{\theta^N})^{-s}\prod_{i=1}^{N}\frac{\Gamma(k+s-1)}{\Gamma(k)}ds\nonumber\\
        &=(\frac{1}{\theta})^N\prod_{i=1}^{N}\frac{1}{\Gamma(k)}\MeijerG[\big]{N}{0}{0}{N}{}{k-1}{\frac{z}{\theta^N}}.
    \end{align}\par
    This completes the proof.
\ifCLASSOPTIONcaptionsoff
  \newpage
\fi

\end{document}